\documentclass[a7paper,portrait]{baposter}

\usepackage[font=small,labelfont=bf]{caption} 
\usepackage{booktabs} 
\usepackage{relsize} 

\usepackage{amsmath,amsfonts,amssymb,amsthm} 
\usepackage{eqparbox}

\usepackage{textcomp}
\usepackage{multicol}
\usepackage{wrapfig} 

\graphicspath{{images/}} 

\usepackage{natbib}
\setcitestyle{authoryear,open={(},close={)}}

 \definecolor{bordercol}{RGB}{40,40,40} 
 \definecolor{headercol1}{RGB}{186,215,230} 
 \definecolor{headercol2}{RGB}{120,120,120} 
 \definecolor{headerfontcol}{RGB}{0,0,0} 
 \definecolor{boxcolor}{RGB}{210,235,250} 

\begin{document}
\newcommand{\hi}{H{\sc i}}
\newcommand{\rabell}{R$_{\mathrm A}$}
\newcommand{\por}{$\times$}
\newcommand{\prim}{$^{\prime}$}
\newcommand{\prin}{$^{\prime\prime}$}
\newcommand{\aprox}{${\sim}$}
\renewcommand{\tablename}{Tabla}
\renewcommand{\figurename}{Figura}

\background{ 
\begin{tikzpicture}[remember picture,overlay]
\draw (current page.north west)+(-2em,2em) node[anchor=north west]
{\includegraphics[height=1.1\textheight]{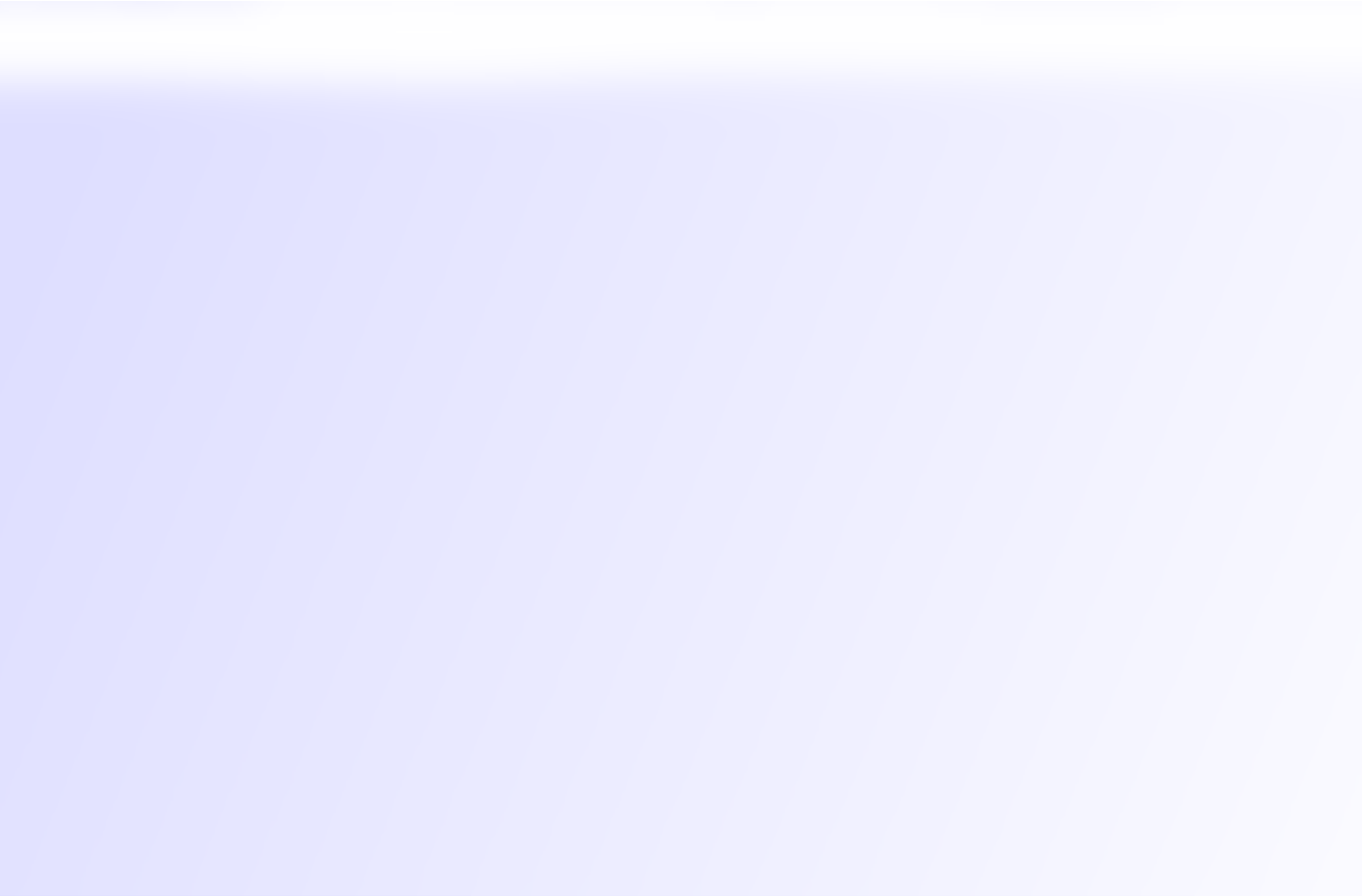}};
\end{tikzpicture}
}

\begin{poster}{
grid=false,
borderColor=bordercol, 
headerColorOne=headercol1, 
headerColorTwo=headercol2, 
headerFontColor=headerfontcol, 
boxColorOne=boxcolor, 
headershape=roundedright, 
headerfont=\Large\sf\bf, 
textborder=rectangle,
background=user,
headerborder=open, 
boxshade=plain
}
{\includegraphics[scale=0.15]{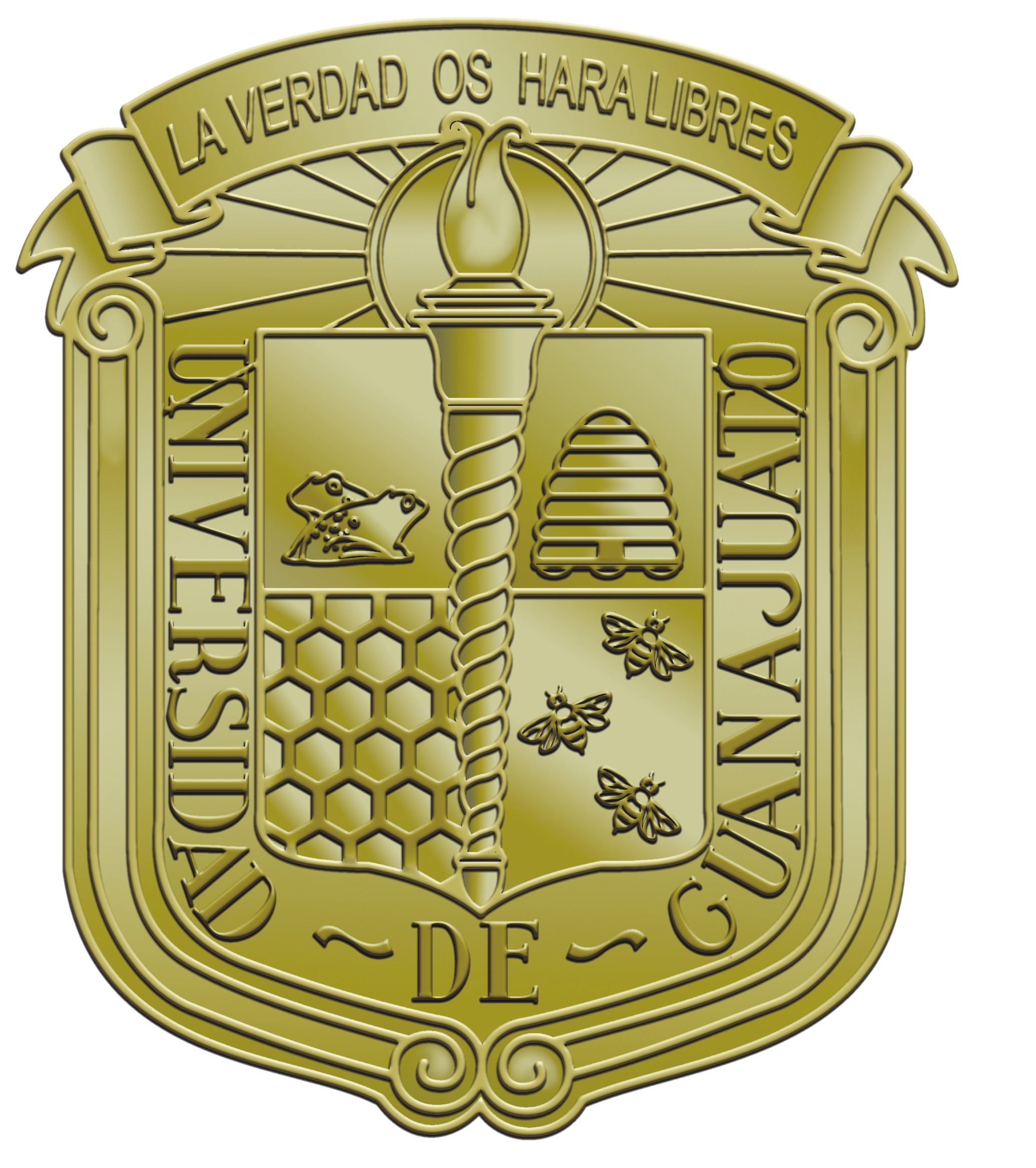}} 
%
%
{ \bf  \LARGE {HI gas, environment, and SF-quenching in the Abell clusters A85/A496/A2670} }
{\vspace{0.3em}
\smaller L\'opez-Guti\'errez, Martha M.,$^1$ Bravo-Alfaro, Hector,$^1$ van Gorkom, Jacqueline H.,$^2$\\Durret, Florence,$^3$ Caretta, C\'esar. A.$^1$\\  

\smaller $^1$\it {Departamento de Astronom\'ia, Universidad de Guanajuato. Mex.} \sf { mm.lopezgutierrez@ugto.mx}\\
 $^2$\it {Department of Astronomy, Columbia University. NY, USA.} 
 $^3$\it {Institut d'Astrophysique de Paris. Paris, Fr.}} 
{\includegraphics[scale=0.2]{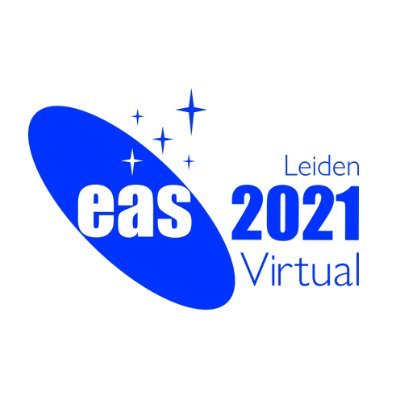}}

\headerbox{Abstract}{name=abstract,column=0,row=0, span=3}{
\small{We study complete sample of spiral galaxies brighter than M$_{\mathrm{B}} \sim$ -18.4 in three Abell clusters: A85 ($z=\,$0.055), A496 ($z=\,$0.033) and A2670 ($z=\,$0.076). This work is based on a large volume limited, blind \hi-survey (21cm, NRAO-VLA), on optical imaging (CFHT) and on the search for dynamical substructures. Our goal is to explore the effects of local environment on the \hi\ gas properties and this, in turn, on the star formation activity and quenching of individual galaxies. We report significant differences among the \hi\ properties of these clusters and we discuss the role played by the large scale structure and by the dynamical evolution of each cluster accounting for the observational evidence.}
}

\headerbox{The Observational Data}{name=observations, column=0,row=0,span=3,below=abstract}
{\small{
We carried out a C-configuration, NRAO-VLA \hi\ survey, with a resolution of \aprox\,24\prin \por \,17\prin, for A85, A496 and A2670 (Table 1).  The surveyed area goes from 2.5 to 4.0 Abell radius for the three clusters (see \textbf{Fig. 1}) and the velocity coverage is larger than three times the cluster velocity dispersion. \hi-mass detection limit is \aprox9.0, 3.0, 8.6$\times$10$^8$\,M$_{\odot}$, for A85, A496, A2670, respectively.  Optical $u'$ $g'$ $r'$ $i'$ images (MegaCam-CFHT) have $1\times 1$~deg$^2$ for the three clusters.  We dispose of redshift catalogues of member galaxies, including 616, 368, 308 objects for A85, A496, A2670, respectively.  By using public databasis (HyperLEDA, NED, SuperCosmos) and carrying out a visual inspection of our RGB frames we built complete catalogues of spirals lying within the \hi\ data cubes and being brighter than M$_{\mathrm{B}} \sim$ -18.4. We report 54, 
32, 46 bright spirals in A85, A496, A2670 respectively.
Hereafter we denote a galaxy as \textit{low mass} if it is less bright than this threshold, equivalent to some 9.3$\times$10$^9$\,M$_{\odot}$.

\begin{multicols}{2}

\begin{center}
        \begin{tabular}{lccc} 
                \hline
                Abell cluster               & A85 & A496 & A2670\\
                \hline
                $\alpha_{2000}$          & 00 41 50 & 04 33 38 & 23 54 14\\
                $\delta_{2000}$          &-09 18 07 &-13 15 33 &-10 25 08 \\
                $z$/$D_C$ (Mpc)          & 0.055/236 & 0.033/141 & 0.076/320\\
                scale 10\prim (Mpc)      & 0.69 & 0.41 & 0.95 \\
                Velocity (km~s$^{-1}$)   & 16,507 & 9,863 & 22,840 \\
                Vel. disp. (km~s$^{-1}$) & 873 & 737 & 745 \\
                R$_{200}$ (Mpc)      & 2.10 & 1.79 & 1.78\\
                1 \rabell (\prim)        & 30.9 & 51.5 & 22.3 \\
                Morph (B-M)          & I & I & I-II \\
                Richness   & 130 & 134 & 224 \\
                $L_X$ ($10^{44}\, \mathrm{erg}^{-1}$)  & 9.4 & 3.8 & 2.3 \\
                LSS                  & SC (11) & isolated & isolated \\
                \hline
                \multicolumn{4}{l}{ }\\
     \end{tabular}
\end{center}
                
\begin{center}
\includegraphics [width=0.31\linewidth] {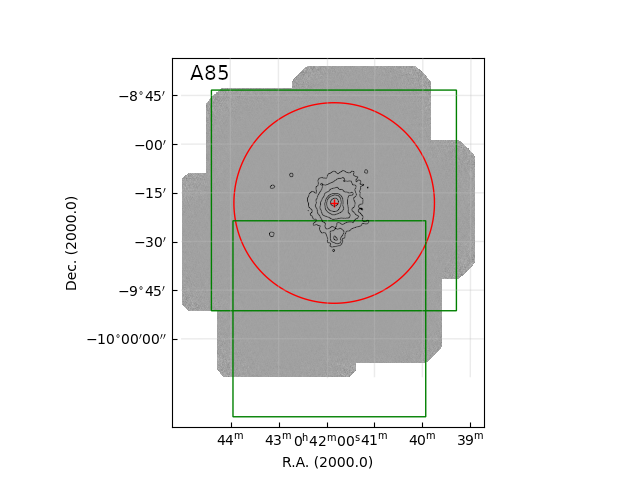}
\hspace{0.12cm}
\includegraphics [width=0.31\linewidth] {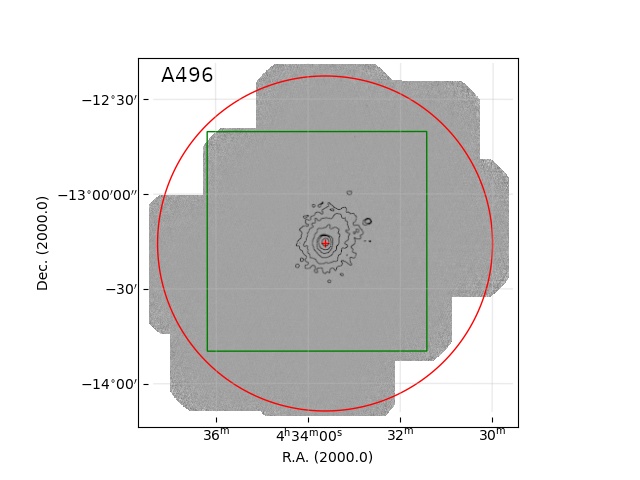}
\hspace{0.12cm}
\includegraphics [width=0.31\linewidth] {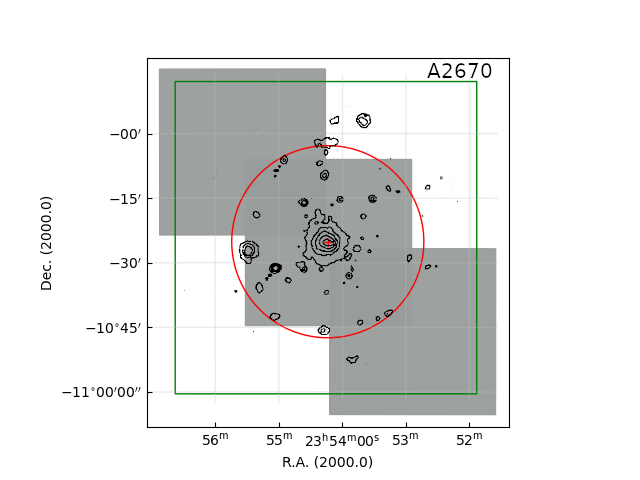}
\end{center}
\small{\textbf{Figure 1.} The \hi\ fields observed with the NRAO-VLA (grey polygons) and the CFHT optical fields (green squares). A85 is in the left panel, A496 in the middle, and A2670 in the right. The \rabell\ is indicated with a red circle and the black contours draw the ROSAT X-ray emission (Durret et al.\,2000). The BCG is indicated with a red cross. }
 
\end{multicols}

}
}
\headerbox{Results}{name=results, column=0,row=0,span=3,below=observations}
{\small{The three clusters have very different \hi\ properties.  \textbf{Fig. 2} displays the distribution of member galaxies (black diamonds); we highlight the bright spirals with blue dots and the \hi-detections with blue  squares. The empty squares correspond to low mass, \hi-rich galaxies. We analyze the distribution of \hi-rich/poor spirals across each cluster, the fraction of low mass galaxies detected in \hi, the number of abnormal \hi\ spirals, etc.  In this work we define a galaxy as $abnormal$ in \hi\, if they are gas deficient or if they display strong asymmetries or offsets, either in position (optical-\hi) or in velocity (optical-\hi).  
We applied two independent methods to trace dynamical substructures (Dressler \& Shectman 1988; Serna \& Gerbal 1996, not shown in this work).   In \textbf{Fig. 3} we show preliminary projected phase-space (PPS, J\'affe et al.\,2015) diagrams, which help to infer the assembly history of a cluster. These plots show the peculiar velocity of each member galaxy as a function of the distance to the cluster center. Symbols for bright spirals and \hi-detections are as in \textbf{Fig. 2}.  Our main observational results are the following (for more details see Bravo-Alfaro et al. 2021, in prep.). \\

{\bf A\,85:} Appears as a typical massive and evolved cluster. We detected only 10 galaxies in \hi, all of them projected to the East, far from the cluster core and  concentrated within a narrow velocity window. No major perturbations are seen in those \hi-rich objects. Two \hi-detections do not correspond to bright spirals; this represents a fraction of 0.2 of low mass \hi-rich objects. Our substructure research confirmed that a number of substructures are undergoing a complex merger process with the main body of A85. \\

{\bf A\,496:}  Previously considered a quasi relaxed system, the \hi\ evidence is very intriguing. First, there is a high number of \hi-detected galaxies, 58, among which one third are classified as $abnormal$. Second, many of these gas rich objects are projected around the cluster center. And last but not least, a striking fraction (equivalent to 0.74) of \hi-detections corresponds to low mass galaxies.   \\

{\bf A\,2670:}  This cluster is dynamically younger than the other two.  We detected 38 galaxies in \hi, packed in two big groups, one is projected onto the core and the other to the SW.  Roughly one third of the gas rich objects are classified as $abnormal$.  A large number (24) of bright spirals are not detected in \hi, but an important fraction (0.42) of \hi-rich galaxies correspond to low mass objects.

}
\begin{center}
\includegraphics [width=0.23\linewidth] {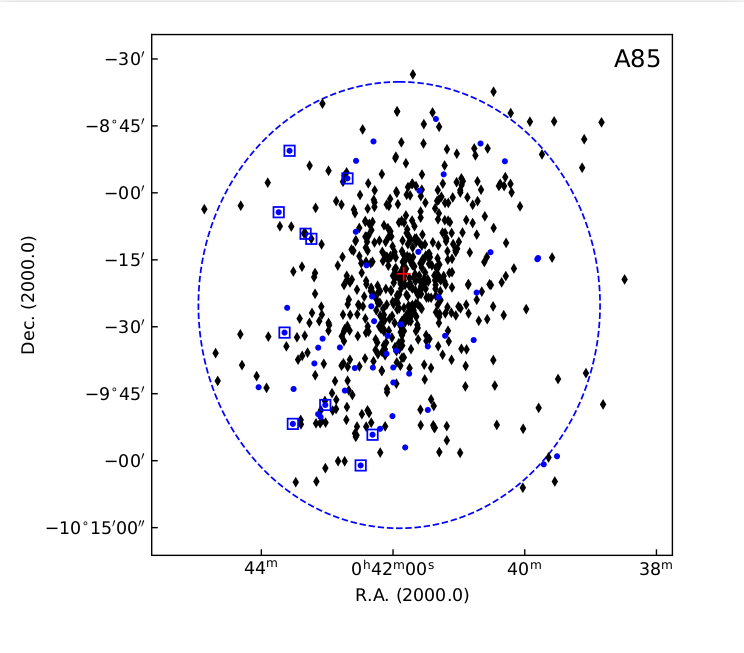}
\hspace{0.12cm}
\includegraphics [width=0.23\linewidth] {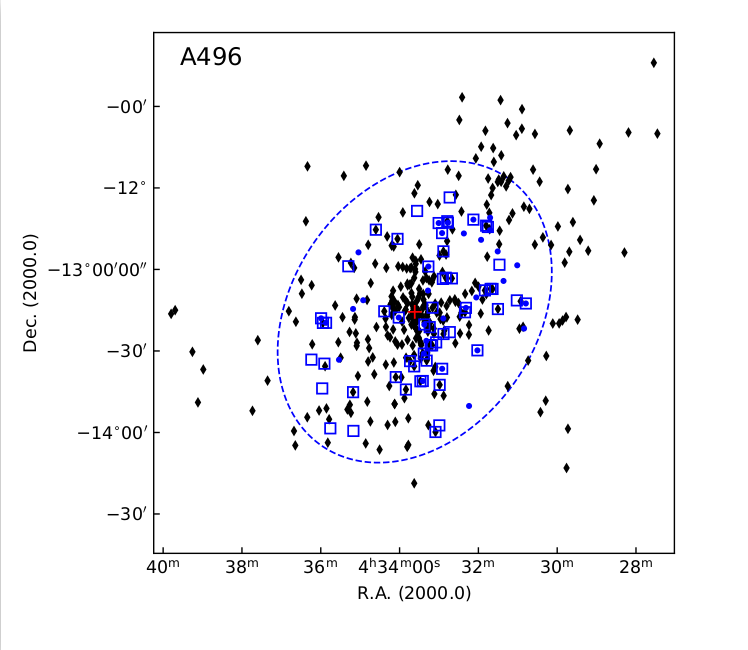}
\hspace{0.12cm}
\includegraphics [width=0.23\linewidth] {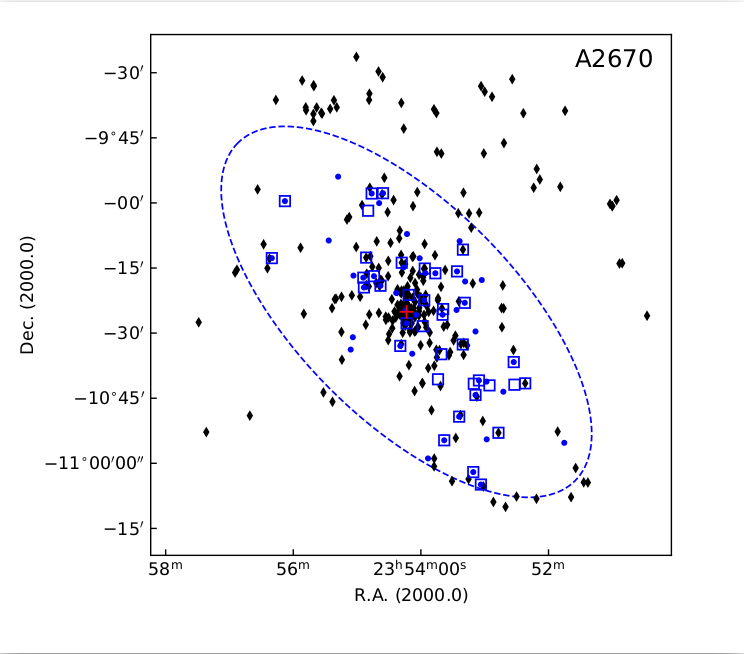}
\end{center}
\textbf{Figure 2.} The distribution of member galaxies in A85 (left), A496 (middle) and A2670 (right). The blue dots are the bright spirals and the blue squares indicate the \hi-detections. The dotted ellipse roughly draws the area observed in \hi\, with the VLA. Each BCG is shown with a red cross.\\\\
\begin{center}
\includegraphics [width=0.23\linewidth] {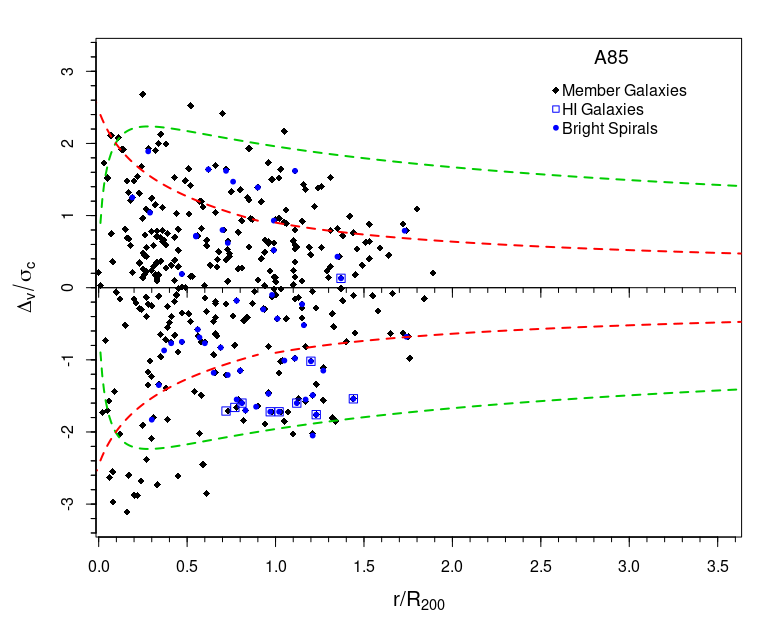}
\hspace{.15cm}
\includegraphics [width=0.23\linewidth] {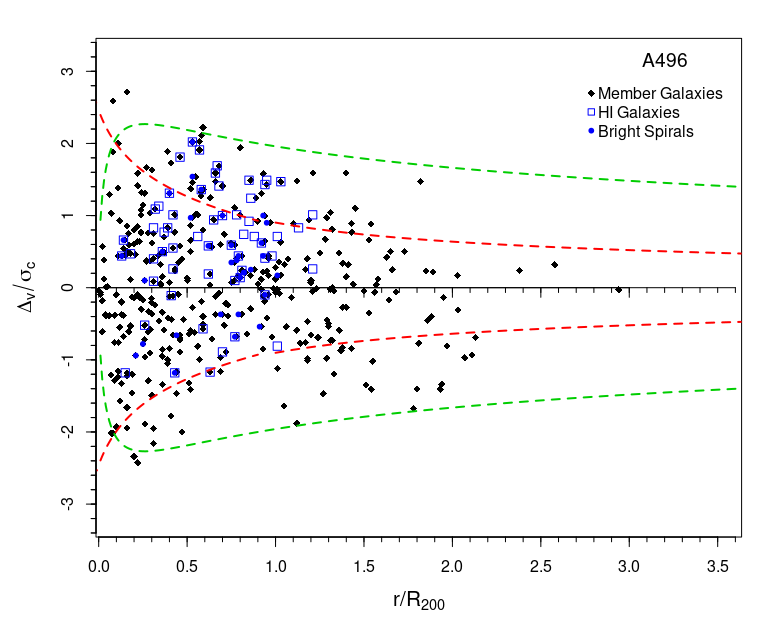}
\hspace{.15cm}
\includegraphics [width=0.23\linewidth] {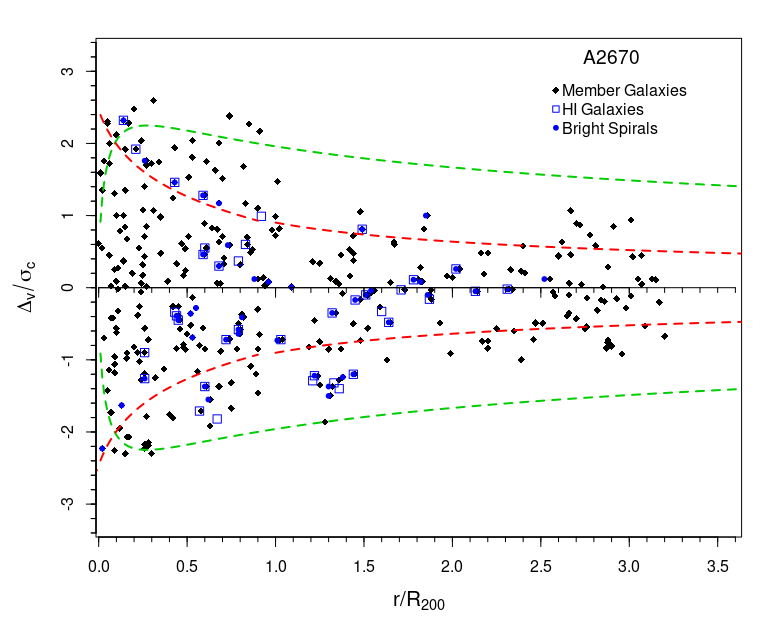}
\end{center}
\small{\textbf{Figure 3.} The projected phase-space diagrams for A85 (left), A496 (middle) and A2670 (right).  The peculiar velocities of member galaxies are normalized by the velocity dispersion of the cluster, and the cluster-centric distance is normalized by R$_{200}$ (following Finn et al.\,2005).  Infall velocities are shown with dashed green lines (following Barsanti et al.\,1981) and the escape velocity is indicated by the dashed red line (following J\'affe et al.\,2015).  We highlight the bright spirals (blue dots) and the \hi-detections (blue squares). }

}

\headerbox{Discussion and Conclusions}{name=conclusion,column=0,below=results,span=3}
{\small{ \hi\ synthesis imaging of large cluster volumes at low redshift, combined with data from other frequencies and with substructure analysis, constitutes a powerful tool to study the evolution of galaxies infalling to massive clusters. \\

\textbf{A85} environment appears to be extremely harsh and similar to the well known case of Coma (Bravo-Alfaro et al. 2000). In A85 all the spirals lying within 1.5 Mpc from the cluster center suffered the stripping of a large fraction of gas (see \textbf{Fig. 2}). On the other hand, the few \hi-rich objects are seen far (in position and velocity) from the cluster center (\textbf{Fig. 3}) and could share a common origin as they have a very low velocity dispersion. These galaxies must be in an early infalling stage, lying at the edge of the infalling velocity limit shown in \textbf{Fig. 3}; this would account for their normal \hi\ appearance. \\

\textbf{A496} has a large number of gas rich, low mass galaxies; 43 of the 58 detections are less bright than the threshold of M$_{\mathrm{B}} \sim$ -18.4. Most of them are located at high velocity, low radius regime (Fig. 3). The low mass, \hi\ rich objects (empty squares) roughly follow the distribution of massive spirals (blue dots and filled squares).  Interestingly, many of these objects are projected around the cluster core within the escape velocity curve, something unexpected for dwarf galaxies which lose their gas more easily than giant spirals.  Projection effects could solve this paradox.  This is supported by the fact that 75\% of all \hi-detected galaxies have radial velocities significantly larger than the cluster velocity. We report that 30\% of detected galaxies are $abnormal$ in \hi, indicating that in spite of possible projection effects, some galaxies are already undergoing strong physical transformations.   \\

\textbf{A2670} was expected to be richer in \hi\ than the other two systems but it has less detections than A496 (38 $vs.$ 58) and it harbors a much lower fraction of low mass \hi-rich objects.  The distribution of bright spirals in A2670 is very complex, with much less LTGs projected near the origin (\textbf{Fig. 3}) compared with A496. \\

By quantifying the general properties of \hi-gas of spirals within large cluster volumes, we confirm that local environment plays an important role in galaxy evolution. 
The differences observed in \hi\ can be correlated with the global properties of the clusters.  For instance, A2670 is less evolved dynamically than the other two systems, but we report less \hi\ detections than in the more relaxed A496. 
A possible explanation solving this paradox is that A496 is embedded in a very long filament (Durret 2021, priv. comm) while A2670 seems to be a rather isolated system. The large scale structure surrounding A85 and A496 would be feeding these systems with a fresh population of LTGs. This process seems to be more advanced in A85 than in A496. 
The present study of \hi\ in individual objects enabled to discover a number of very perturbed galaxies in these clusters. In a forthcoming work we will compare our multi-frequency observations with hydrodynamic models in order to determine the specific mechanisms driving galaxy transformation under different cluster environments (L\'opez-Guti\'errez et al., 2021, in prep.)  Individual \hi-maps and velocity fields for all spatially resolved galaxies in these three clusters will be available in Bravo-Alfaro et al. 2021, in prep.
}
}
\headerbox{References}{name=references,column=0,below=conclusion,span=3}{\small

Barsanti S., 
et al., 2018, ApJ, 857, 71\\
Bravo-Alfaro H., 
et al., 2000, AJ, 119, 580\\
Dressler A., \& Shectman S.~A., 1988, AJ, 95, 985\\
Durret F., 
et al., 2000, A\&A, 356, 815\\
Finn R.~A.,
et al., 2005, ApJ, 630, 206\\
Jaff{\'e} Y.~L., 
et al., 2015, MNRAS, 448, 1715\\
Serna A., \& Gerbal D., 1996, A\&A, 309, 65

}
\end{poster}
\end{document}